# Spin structure of Graphene/Pt interface for spin current formation and induced magnetization in deposited (Ni-Fe)-nanodots

A.M. Shikin[1*], A.A. Rybkina[1], A.G. Rybkin[1], I.I. Klimovskikh[1], P.N. Skirdkov[2,3], K.A. Zvezdin[2,3], A.K. Zvezdin[2,3]

[1] Saint Petersburg State University, Saint Petersburg, Peterhof, Ulyanovskaya str. 1, 198504 Russia

[2] A.M. Prokhorov General Physics Institute, Russian Academy of Sciences, Moscow, Vavilova str. 38, 119991 Russia

[3] Moscow Institute of Physics and Technology, Dolgoprudny, Institutskiy per. 9, 141700 Russia

**Abstract**

Spin electronic structure of graphene π-states and Pt 5d-states for the Graphene/Pt interface has been investigated. Here, we report a large induced spin-orbit splitting (~70-100 meV) of graphene π-states with formation of non-degenerated Dirac-cone spin states at the K-point of the BZ crossed with spin-polarized Pt 5d-states at Fermi level that opens up a possibility for creation of new spintronics devices. We propose to use this spin structure for formation of spin current with spin locked perpendicular to the momentum for induced remagnetization of the (Ni-Fe)-nanodots arranged atop the interface. Theoretical estimations of the spin current created at the Graphene/Pt interface and the induced intrinsic effective magnetic field leading to the in-plane remagnetization of the NiFe-nanodots due to spin-orbit torque effect are presented. By micromagnetic modeling based on experimentally observed spin-orbit splitting we demonstarte that the induced intrinsic magnetic field might be effectively used for magnetization swithching of the deposited (Ni-Fe)-nanodots.

**Main**

Controlable manipulation of magnetization in low-dimensional systems without using external magnetic field using the combination of the exchange magnetic and spin-orbit interaction originated from analysis of the features of spin electronic structure attract an enhanced attention last time due to a high demand of a new generation of the energy efficient memory and quantum logic devices [1-4]. Recently, a very perspective idea was proposed [5-7] and then experimentally confirmed [8-10] related to the use of the spin current created in low-dimensional Rashba systems with strong spin-orbit interaction for the control of the magnetization state in ferromagnetic metals (FM) attached to these Rashba system by the induced spin-orbit torque effect. In Refs. [8,9] this idea was applied to the AlOx/Co/Pt system where corresponding spin current was formed due to structural inversion asymmetry at the borders of the Co-film. This current induces an effective intrinsic transverse magnetic filed which switches the initial out-of-plane magnetization in the Co-film due to the spin-orbit torque effect [5-7,11,12]. For electron moving in two-dimensional system with out-of-plane potential gradient (caused by the asymmetric crystal field potential noted above) the spin-orbit coupling leads to formation of an effective magnetic field in electron's frame. The Rashba spin-orbit interaction produces a non-equilibrium spin density proportional to the current density which interacts with the magnetization via the exchange coupling and ultimative rotates the initial magnetization due to spin torque effect [5-7]. Thereat, the reversing of the current applied to the system is followed by the remagnetization of the ferromagnetic (Co) film. We have to note than an alternative idea which descibes the observed magnetization switching in the systems with strong spin-orbit interaction by the spin torque induced by the spin Hall effect in the system with strong spin-orbit interaction is also widely discussed in literature [13-15], see discussion below.

The influence of the current-driven spin-orbit torque induced by the Rashba effect on the magnetization switching the dynamics is estimated to be even more effective than the manipulation by external magnetic field [8,9,14]. Therefore, a combination of the spin-orbit Rashba interaction and the spin-orbit torque effect open the perspectives for designing of the spintronics devices with reduced dimensions and energy consumption. Thereat, a search of new perspective alternative systems with the features of spin structure allowing switching magnetization by the spin-orbit torque with new proposals of corresponding spin current formation and detailed investigations of their spin structure are very actual and important.

Here, we report the details of the spin structure of the Graphene/Pt interface which we propose to use as an effective source for the spin current formation under an electrical or thermal gradients applied. The spin-orbit interaction at the Graphene/Pt interface can produce non-equilibrium local spin density which may induce the magnetization dynamics in FM nanodots arranged atop the interface. We demonstrate that the graphene/Pt interface is characterized by an enhanced spin-orbit splitting of the graphene $\pi$ states with the in-plane orientation of spin strongly locked perpendicular to the momentum with formation of non-degenerated Dirac-cone spin states near the Fermi level. Basing on the experimental investigation of the spin structure of the Graphene/Pt interface we propose the idea and schematic construction how this unique spin structure of the graphene-derived system can be used for formation of the spin current and corresponding magnetization switching in ferromagnetic (Ni-Fe)-nanodots arranged atop the interface due to the developed spin-orbit torque effect. In our knowledge it is the first time proposal of using of a graphene-derived system as an active element for switching the magnetization in the attached FMs.

Generally speaking, for effective formation of highly-polarized in-plane spin current caused by the spin-orbit interaction the use of the systems with non-degenerated Dirac-cone spin structure is the most optimal. Electronic structure of graphene is characterized by the spin-degenerated Dirac-cone structure with linear dispersion that determines the zero effective mass of the Dirac fermions, a high Fermi velocity, anomalously large carrier mobility in graphene [16] leading to its unique physical and transport properties including a high length of spin relaxation. Unfortunately, a very small magnitude of the spin-orbit splitting of the graphene π states (less that 1 meV [17]) does not allow to use it directly for formation of the spin current with effective spin polarization. However, a possibility of induced anomalously large spin-orbit splitting for the states in light metals (Ag,Cu,Al) was demonstrated recently [18-22] when thin films of the light metals were deposited on surfaces of heavy metals, for instance, W(110). Similar effect has been demonstrated also for graphene [23-25] brought in contact with heavy metal Au revealing a noticeable induced spin-orbit splitting of graphene π-states near the Fermi level. It was shown that the effect of the enhanced spin-orbit splitting of the graphene π states is caused by the hybridization between the π states of graphene and Au 5d states due to corresponding spin-dependent avoided-crossing effects. Thereat, in the region of the linearity of the π state dispersion near the K point of the graphene Brillouin Zone (BZ) near the Fermi level the spin-orbit splitting reaches the value of about 100 meV.

In the current work we present the features of spin electronic structure of the Graphene/Pt interface, i.e. for graphene contacting with other heavy metal (Pt) which is one of the mostly spreading metals in spintronics. Pt is effectively used in spintronics both as a spin current absorber and a spin current source. Pt is characterized by the spin-Hall conductivity which is significantly larger than that in semiconductors and other metals [26-28]. Pt is widely and successfully used in numerous experiments related to the influence of the formed spin current on induced remagnetization in FMs contacting with Pt layers [8-10,13-15, 26,29-32]. It is also used as an important element in constructions related to investigations of the spin torque effect and for registration of the developed spin current including the case of the spin Seebeck effect [33,34]. The high value of spin Hall conductivity in Pt is attributed both to a high spin-orbit coupling and a large number of carriers caused by a high degeneracy of the 5d spin-polazized states near the Fermi level. Among metallic systems Pt shows remarkably large spin Hall effect (larger that in Cu and Al) [27,28]. For the Graphene/Pt interface the high density of spin-polarized Pt 5d states is confirmed by analysis of the spin structure near the Fermi level carried out in the present work. In our work we have shown that the spin structure of the graphene π-states is characterized by the Dirac-cone states with localization of the Dirac-point in the region of the Fermi level with the value of the induced spin-orbit splitting of the π-states ~70-100 meV near the Fermi level. These spin-polarized π states are crossed with the spin-polarized Pt 5d states at the Fermi level that produces a basis for the spin injection between the states and an opportunity of developing effective spin Hall effect. We propose the idea how this spin structure can be used for creation of electrically- or thermally-driven spin current accompanied by the induced remagnetization of array of FM-nanodots deposited atop the Graphene/Pt interfaces. For formation of the FM-electrodes we propose to use the permalloy (Ni-Fe) nanodots (with stoichiometry $Ni_{81}Fe_{19}$) and Au monolayer between the (Ni-Fe) and graphene that prevents to the distortion of graphene Dirac-cone states under contact with (Ni-Fe), see, for comparison, [23-25,35-37]. Thereat, the Graphene/Au/(Ni-Fe) interface is also characterized by a large spin-orbit splitting of the graphene π states and the Dirac-point position practically arranged at the Fermi level [23,25] that supports formation of a necessary spin current at the interface. We have described in details the features of the spin structure of all principal elements included in the proposed construction and have presented theoretical estimations of the spin current formed at the Graphene/Pt interface due to application of an electrical or thermal gradient and have calculated the induced spin-orbit torque leading to remagnetization of the (Ni-Fe)-nanodots. It was shown by the micromagnetic modeling that the magnetic field which is formed due to experimentally observed spin-oirbit splitting is enough for the remagnetization of the (Ni-Fe)-nanodots with the typical sizes.

**Experimental Results and Discussions**

Fig.1.a plots the dispersions of the graphene π-states and Pt 5d states measured in the ΓK-direction of the BZ for graphene synthesized by cracking of propylene on Pt(111), see details of the system preparation in Methods. The dispersions are presented for the part of the BZ for the binding energies between 2.5 eV and the Fermi level. Here, the positions of the K-point of the graphene BZ located at $k_{\parallel}$ = 1.7 Å$^{-1}$ and of the M-point of the Pt(111) BZ located at $k_{\parallel}$ = 1.3 Å$^{-1}$ are noted. The schematic images of the graphene and Pt(111) BZs are shown in the inset in Fig.1.a by white solid and dotted lines, respectively. The BZs are rotated on 30° relative to each other. The presented dispersions were constructed with use of the experimental valence-band spectra measured by angle-resolved photoemission at a photon energy $h\nu$ = 62 eV with p-polarization of the incident synchrotron radiation.

The branch of the graphene π states reaches the Fermi level and cross it at the K-point of the graphene BZ passing from the first to the second BZ. For more clear presentation of the character of the graphene π state dispersion relations in Fig. 1.b the dispersion measured in the direction through the K-point, but perpendicular to the ΓK-direction is presented. This direction is shown by white arrors in the inset in Fig. 1.b. This dispersion is measured without transition between the first and the second BZs. Therefore, due to similarity in the matrix elements on both sides relative to the K-point, the measured dispersion has more symmetric character in comparison with that in Fig.1.a, where practically only one branch in the first BZ is visible. Fig. 1.b shows pronouncedly linear character of the π state dispersion near the K-point of the BZ for the Graphene/Pt interface. The Dirac point corresponding to the crossing of the Dirac-cones of the filled and empty states lies practically at the Fermi level. The Pt 5d states are characterized by series of the branches crossing the Fermi level in the region of the M-point of the Pt(111) BZ. They are shifting above the Fermi level with $k_{II}$ and are coming back with crossing the Fermi level in the second BZ of Pt(111) near the K-point of the graphene BZ. In the region of the K-point of the graphene BZ the branches of the Pt 5d states cross the branch of the graphene π states at the energy of 0.7-1.0 eV and near the Fermi level. In the region of the crossing between the graphene and Pt states the distortions and formation of local breaks in the dispersion of the π states are observed. It is related to the hybridization between the Pt 5d and graphene π states and can be described by corresponding spin-dependent avoided-crossing effect which modifies significantly the spin structure of the graphene π states. The distortion of the graphene π states under crossing with the Pt 5d states is pronouncedly visible in the photoemission spectra measured with opposite elliptic polarization of the incident synchrotron radiation. The dispersions measured by circular dichroism with positive (blue) and negative (red) elliptic polarizations are shown in Fig. 2.a. The splitting of the branches is related to different sensivity of the states with different spin orientation to the photoexcitation with opposite elliptic polarization of light. Fig. 2.b shows a set of the spin- and angle-resolved photoemission spectra measured at different values of $k_{II}$ ($k_x$) in the region of the K-point of the graphene BZ. The values of $k_{II}$ corresponding to the spin-resolved spectra in Fig. 2.b are shown in Fig. 2.a by dashed lines. Analysing the spin-resolved spectra in Fig. 2.b we can see that the Pt 5d states are well spin-split. Just the effect of the hybridization of the graphene π states with the spin-polarized Pt 5d states leads to the "induced" spin splitting of the π states, similar to that as in the case of the Graphene/Au interface [23,25]. The corresponding spin splitting of the graphene and Pt 5d states is shown in Fig. 2.b by red and blue vertical lines. One can see that the graphene π states are really spin-split due to the spin-depedent hybridization with spin-polarized Pt 5d states. The value of the spin splitting of the graphene π states close to the K-point is of about 70-100 meV at the Fermi level in different directions of the BZ. This spin spitting is better visible in Fig. 2.c where the spin resolved spectrum is presented for $k_{II}$ ($k_y$) close to the K-point of the BZ. This spectrum was measured in the geometry as used for the measurement presented in Fig. 1.b, where the contribution of the graphene π states was higher than that of the Pt 5d states. To separate the contributions of the graphene π and Pt 5d states comparative measurements with the linear p- and s- polarization of the incident synchrotron radiation were carried out. The dispersion measured with the p-polarization (sensitive to the π states of graphene) is presented in Fig.1.b. The dispersion measured with the s-polarization is presented in Fig.1.c. This geometry of the experiment is insensitive to the π states of graphene and, therefore, the dispersion in Fig.

1.c shows mainly the contribution of the Pt 5d states. By this separation of the contribution of the graphene π and Pt 5d states we confirm that the spin-polarized graphene π states are crossed with the spin-polarized Pt 5d states at the Fermi level in the region of the K-poin of the BZ. It is very important for a possibility of application of this system in spintronics. Firstly, the value of the spin splitting of the graphene π states at the Fermi level is already enough for effective creation of spin current. Secondly, the intersection between the spin-polarized graphene and Pt states at the Fermi level can produce a possibility of effective spin injection between the Pt 5d and graphene π states and corresponding spin transport that is important for the spin current formation and corresponding induced spin torque effect. Third, the spin orientation at the Graphene/Pt interface is strongly locked perpendicular to the electron momentum as for two-dimensional system with enhanced spin-orbit coupling that provides a good basis for a high selectivity of spins in the forming spin current. Additionally, such spin structure of Pt d- and graphene π- states can lead to a possibility of high value of spin Hall conductivity.

Fig. 3.a shows schematically the spin structure energy diagram for the Graphene/Pt interface with the Dirac-cone states located at the K- and K'- points at $k_{II} = +1.7$ and $-1.7$ Å$^{-1}$, respectively. Due to enhanced spin-orbit splitting the unique spin structure is formed near the Fermi level in the energy range corresponding to the π-state spin splitting (~70-100 meV below and above the Fermi level). This region is already characterized by the single branches of non-degenerated Dirac-cone spin states with linear dispersions and the in-plane spin orientation locked to the momentum of electron. If to apply an electrical or thermal gradient along the interface the electrons from occupied (π) states will be excited to unoccupied (π*) states of graphene near the Fermi level, see the bent black arrows in Fig. 3.a. Other part of electrons can be injected into the unoccupied π* states of graphene from the spin-polarized Pt 5d states with the same spin polarization. As a result, in the direction of applied electrical or thermal gradient (for instance, along $k_x$), due to an addition in momentum ($+\Delta k_x$) the contribution in the forming current of electrons in the ($+k_x$)-direction will be larger than that in the ($-k_x$)-direction, see the diagram in Fig. 3.b. Taking into account that in graphene spin is strongly locked perpendicular to the momentum this difference between the oppositely-oriented contributions produces an uncompensated spin current at the Graphene/Pt interface in the direction of the formed electrical field with the spin oriented perpendicular to the momentum. As we noted earlier this spin current can induce a magnetization in FM-nanodots deposited atop the Graphene/Pt interface. Thereat, the charge current flowing through a bulk of platinum parallel to the interface creates the spin current penetrating into FM in the $k_y$-direction due to spin Hall effect with the same spin orientation leading to the same direction of the induced magnetization in FM atop [9,13-15,26]..

If to change the electrical field or thermogradient on opposite ones the direction of the spin current and its spin polarization will be changed on opposite one, too.

For realization of the idea of the magnetization induced in the FM-nanodots arranged atop the Graphene/Pt interface by the spin current developed at the interface one can use the classical construction proposed in Refs [8-10,13-15,31,32], only with the principal differences in the origin of the spin polarization and source of the spin current. Moreover, in opposite to the idea developed in Refs [8-10] with the spin current formed in the Co-film due to the Rashba effect leading to the out-of-plane magnetization we propose to use the graphene-derived system characterized by the non-degenerated Dirac-cone spin structure with the in-plane spin orientation and corresponding in-plane magnetization of FM. A structural cell based on the Graphene/Pt interface which we propose to use is shown schematically in Fig. 3.b. An electrical or thermal gradient applied to the Graphene/Pt stripes creates a spin current with the in-plane orientation of spin locked perpendicular to the moving electrons that induces corresponding in-plane magnetization in permaloy ($Ni_{81}Fe_{19}$)-nanodots due to spin-orbit torque effect.

The system can be formed on the SiC-substrate by deposition of Pt-layer with consequent synthesis of graphene or by intercalation of Pt underneath a graphene on SiC, see details in Methods.

**Spin current development**

Let's consider the formation of the spin current at the Graphene/Pt interface due to application of an electrical or thermal gradient. For description of electronic spin transport in graphene layer with strong spin-orbit interaction we use the Boltzmann equation in the time-relaxation approximation. Application of electric field $E_x$ along x-direction causes a spin current alone the x-axis:

$$j_S^x = \frac{2\alpha}{\hbar \upsilon} j_e^x = \frac{2\alpha}{\hbar \upsilon} \sigma E_x \quad (1),$$

where $\sigma$ is the graphene conductivity, $\alpha$ is the constant of spin-orbit (Rashba) interaction ($\alpha \cdot k_F = 80$ meV) with Hamiltonian $H_{SO} = \varepsilon(k_x) - \frac{\alpha}{\hbar}(k_x \sigma_y - k_y \sigma_x)$. Here $\varepsilon(k_x) = \hbar \upsilon k_x$ is the dispersion law for the electrons of graphene. In the case of the applied thermogradient $\partial T/\partial x$ along x-direction the spin current along the x-axis can be expressed by the Eq.1 after the replacement $\sigma E_x \to -S \cdot \partial T/\partial x$, where $S$ is the Seebeck constant. For both cases this leads to spin accumulation: appearance of uncompensated spin density with the spin polarization in the y-direction (the spin is locked perpendicular to the momentum):

$$\langle \delta \vec{\sigma} \rangle_y^{\nabla T} = -\frac{\alpha}{e \hbar \upsilon^2} S \frac{\partial T}{\partial x} \quad (2),$$

$$\langle\delta\vec{\sigma}\rangle_y^{\nabla U} = -\alpha\frac{ne\tau}{\hbar\varepsilon_F}E_x \quad (3),$$

where $n$ is the electron density in graphene, $\tau$ is the relaxation time. Taking into account that the velocity of electrons at the Dirac point is $\upsilon = 10^6$ m/s [39] and using Eq.1-2 we can estimate the dimensionless ratio of the spin and the electric currents in both cases: $|j_S|/|j_e| = 0.015$. At the same time in conventional generators of the spin current, based on the spin-Hall effect (SHE), this ratio is of the order of 0.07 [27,28] or 0.03 [31]. Hence, we have comparable efficiency of the spin current excited at the Grapene/Pt interface due to applied electrical or thermogradient and, for instance, in Pt [13-15,26-28] due to SHE. However, unlike the SHE case, in the investigated system the direction of the spin current and electric currents coincides. Using the values of the Seebeck constant $S^* \sim 30$ mV/K [38] and graphene conductivity up to $\sigma \sim 2\times 10^3$ S/cm [54], we can estimate the value of an electric current induced by the temperature gradient: $j_e^x \approx 6\cdot 10^5 \left(\frac{\Delta T[mK]}{\Delta X[nm]}\right)$ A/cm$^2$, where $\Delta T$ is the temperature difference between the two contacts in millikelvins, $\Delta X$ is the distance between contacts in nanometers. In the case of applied electrical gradient electric current can be estimated as follows: $j_e^x \approx 2\cdot 10^7 \left(\frac{\Delta U[mV]}{\Delta X[nm]}\right)$ A/cm$^2$, where $\Delta U$ is the the electrical potential difference between the two contacts in millivolts.

According to this estimation we can say that electric, and as a consequence, spin current developed at the Graphene/Pt interface due to application of electrical or thermogradient, at least, is not less than that in the case of the typical SHE. Thus, the Graphene/Pt interface characterized by unique spin structure (shown schematically in Fig. 3.*a*) can be effectively used for excitation of a spin current under application of an electrical or thermal gradient, i.e. can play a role of effective generator of spin currents with the in-plane spin orientation strongly locked perpendicular to the momentum of the moving electron.

**Application of the Graphene/Pt interface for the magnetization dynamics of the (Ni-Fe)-nanodots array due to the spin-orbit torque effect**

As we noted above, the spin orientation in the current formed at the Graphene/Pt interface can play a role of effective intrinsic magnetic field induced by the Rashba effect which can lead due to the exchange interaction to the induced spin polarization of the conductivity electrons followed by the induced remagnetization of the FM-nanodots arranged atop the Graphene/Pt interface. The process of remagnetization of the FM-nanodots should be maximally effective in the case, when the initial direction of the magnetization is parallel to the intrinsic magnetic field induced by the Rashba effect. Therefore, we propose in our construction in Fig. 3.b to use the (Ni-Fe)-nanodots with the magnetization induced parallel to the Graphene/Pt interface similar to that used in Ref [13-15,31.32], in opposite to the perpendicular magnetization of the Co-films used in Ref [8,9]. Permalloy is a ferromagnetic alloy ($Ni_{81}Fe_{19}$) allowing effective remagnetization induced parallel to the surface. It is a soft magnetic, therefore, a lower effective magnetic field is demanded for its remagnetization. (It is 565 Oe for Fe, 7429 Oe for Co and 233 Oe for Ni, for comparison [11]).

So, in the proposed consruction the induced remagnetization is produced by the spin-orbit torque effect (with in-plane magnetization) by the spin current formed in two-dimensional graphene-derived system with in-plane orientation of spin locked perpendicular to the momentum. Thereat, the direction of the spin current and corresponding orientation of spin can be reversed in dependence on the direction of the applied electrical or thermogradient.

As we noted before, the spin-orbit torque effect was intensively studied before for the AlOx/Co/Pt system with the structure inversion asymmetry at the borders of the Co-film [8,9] and it was demonstrated [9,29,30] that through a combination of spin-orbit and the exchange interactions an electric current flowing in the plane of a Co layer with asymmetric Pt and $AlO_X$ interfaces produces an effective transverse magnetic field of 1 T per $10^8$ A/cm$^2$. Through the s-d exchange coupling this effect produces an opportunity to manipulate the magnetization. Remagnetization of the Co-film can be carried out by domain-wall (DW) motion excitation [29,40]. Important, that in this case switching of the magnetization or DW-excitation is achieved using only the spin-orbit field, without any extra external fields. On the other hand, there is a number of works which relate the formed magnetization to the spin Hall effect (see, for instance, [13-15]). According to these works the intrinsic spin Hall effect taking place for the electrons flowing through Pt converts the applied in-plane charge current in Pt (as in a metal with strong spin-orbit interaction) into a transverse spin current which propogates into FM through the interface. The spin polarization of this spin current due to deflection of electrons with opposite spin in opposite directions is directed perpendicular to the interface. The forming spin polarization induces the corresponding magnetization in FM which is transferring into the component with vertical magnetization in the case of the Pt/Co/AlOx system due to the spin transfer torque effect [8-10]. In the case of Pt/$Ni_{81}Fe_{19}$ system with the in-plane initial magnetization [13-15,31,32] the induced remagnetization is considered as oriented directly along the spin in the injected spin current. It is interesting that the direction of the spin orientation leading to corresponding remagnetization in FM due to the spin Hall effect and the direction of the intrinsic effective magnetic field induced by the Rashba effect are similar. Therefore, at present moment it is difficult to distinguish the influence of the spin-orbit torque induced by the Rashba and the spin Hall effects. While, with decreasing the thickness of the Pt-layer the direct current through the Pt-layer and, consequently, an influence of the spin Hall effect developed in Pt can be significantly reduced. The system with monolayer-like Pt can be formed by an intercalation of Pt underneath a graphene on SiC.

If we consider the Graphene/Pt interface, the spin current created at this interface can be effectively injected into the FM *d*-minority-spin-states due to well correlation between the spin structure of the Graphene/Pt interface and the FM spin-polarized d-states near the Fermi level in the region of the *K*-point of the BZ. A possibility of injection of spin-oriented electrons between the graphene π-states and the FM *d*-minority-spin-states is discussed, in particular, in [36,41,42]. This idea underlies the operation of graphene spin filter (see, e.g., [36,43]). Injection of electrons between these states is possible due to crossing of the FM *d*-minority-spin-state branch with that of graphene π-state in the region of the Fermi level. The FM *d*-majority-spin-states have a binding energy ~1 eV and practically should not contribute to the process. Injection of the spin-oriented electrons from Graphene/Pt into the FM *d*-minority-spin-states induces the corresponding spin moment in the FM-layer. A possibility of spin injection from graphene or through graphene into FM with effective remagnetization of FM-electrode was also discussed in literature, for instance, in Ref. [44] for the Co/Graphene/FeNi system. Analogously, an effective spin injection from ferromagnetic $Ni_{80}Fe_{20}$-electrode into silicon through graphene [45] was experimentally observed. As to influence of the Au interlayer used for preserving the Dirac-cone structure of graphene (see Fif. 3.b), in Ref. [46] it was noted that Cu-interlayer between (Ni-Fe) and graphene does't reduce the spin injection, at least. While in Ref. [26] an exponential decreasing of effective induced magnetization through the Cu-wire with increase of the distance was observed. It means that the Au monolayer introduced between the Graphene/Pt interface and (Ni-Fe) in Fig. 3.b should not decrease significantly the effeciency of remagnetization of (Ni-Fe) induced by the spin current developed at Graphene/Pt interface. Thereat, a graphene, too, as an effectively tunnel barrier will not restrict a transfer of spin current into (Ni-Fe).

For analysis of possibility of magnetization of the (Ni-Fe)-nanodots deposited on the Graphene/Pt interface let's consider an emergence of the effective spin-orbit induced magnetic field in the magnetic (Ni-Fe) film induced by thermally or electrically driven spin current due to the exchange interaction between the π-electrons of graphene (under contact with Pt) and the local magnetic moment of the magnetic layer. The effective induced magnetic field $\vec{B}_{SO}$ can be represented in following form:

$$\vec{B}_{SO} = -\frac{\alpha k_F}{e \upsilon M_S} P \cdot j_e \left[ \vec{z}^{\,0} \times \vec{j}^{\,0} \right], \quad \text{where} \quad j^0 = 1, \quad P = \frac{J}{\varepsilon_F}. \tag{4}$$

This additional effective field leads to appearance of the spin torque, acting on the magnetization in the (Ni-Fe) layer. Corresponding spin torque per unit volume is given by:

$$\vec{T}_{SO} = \left[\vec{M} \times \vec{B}_{SO}\right] = \frac{\alpha k_F}{e\upsilon M_S} Pj_e \left[\vec{M} \times \vec{j}^0\right] = \frac{\alpha k_F}{e\upsilon} Pj_e \left[\vec{M}^0 \times \vec{j}^0\right] \qquad (5),$$

Now the effects of transfer of the domain wall along ferromagnetic wires under influence of spin-polarized current pulses is intensively studying [2,5-15, 47-51] and is considered as a basis for the magnetic domain-wall racetrack memory [2,4,48]. However, for the practical use of such effects, it is required to improve the efficiency of action of the current on a domain wall. Recent theoretical [52] and experimental [53] studies, based on perpendicular current junction, demonstrate rather high velocities (up to 500 m/s) and relatively low current densities, required for an efficient DW's dynamics excitation. At the same time, it was demonstrated in Ref. [51] that the induced spin-orbit torque has similar behavior and can lead to the efficiency of the DW motion comparable with the aforementioned spin transfer torque. These results allow us to expect high efficiency of the spin-orbit torque, caused by the presence of an electrical or temperature gradient. To demonstrate it we performed the series of micromagnetic simulations using the experimentally observed spin-orbit splitting value (see details in Methods). The hysteresis loops for the (Ni-Fe)-nanodot with the size $50 \times 50 \times 2$ nm$^3$ are represented in Fig.4.a for different value of the anisotropy constant. As one can see from these results, even in the case of additional anisotropy (intrinsic bulk anisotropy of permalloy is almost zero), which usually added for the stabilization of two opposite magnetization directions, the critical switching field is not more than 90 *Oe* (corresponds to $1.8 \times 10^7$ A/cm$^2$). In the case of only shape anisotropy (K=0) the critical switching field is not more than 20 *Oe* (corresponds to $4 \times 10^6$ A/cm$^2$). To produce this effective field it is enough to apply temperature gradient of about 6.7 mK/nm or electrical potential difference of about 0.2 mV/nm between the two contacts. At the same time, the critical remagnetization field is essentially independent on the size of the FM-nanodot for hard magnets ($K \neq 0$) and only slightly increases with the size for soft magnets (see Fig.4.b), since in this case it is determined by the shape anisotropy. This proves the possibility of an effective remagnetization of the FM-nanodots with different sizes and different anisotropy using the effective magnetic field formed by the spin current at the Graphene/Pt interface due to Rashba effect.

In summary, we have demonstrated effect of the enhanced spin-orbit splitting of the π states of graphene of 70-100 meV near the Fermi level evolving under interaction with Pt. The spin-polarized graphene π states and Pt 5d states are crossed at the Fermi level in the region of the K-point of the BZ that produces a basis for the spin injection between the states and developing effective spin transport and spin Hall conductivity. Basing on the experimentally measured spin structure of the Graphene/Pt interface with the non-degenerated Dirac-cone spin states near the Fermi level we proposed to use it for formation of the spin current with the in-plane spin orientation strongly locked perpendicular to the momentum. This spin current can lead to the induced in-plane remagnetization of the (Ni-Fe)-nanodots arranged atop the interface due to the spin-orbit torque effect. We have presented theoretical estimations of the spin current developed at the Graphene/Pt interface by application of an electrical or thermogradient and the efficiency of corresponding induced spin-orbit torque. By micromagnetic modeling we have demonstrated that for the observed spin-splitting value the formed intrinsic magnetic field is enough for remagnetization of the deposited NiFe-nanodots.

**Methods**

The presented and discussed systems are formed by the following operations. Graphene monolayer is synthesized by cracking of propylene, $C_3H_6$, (or other carbon-contained gases) at surfaces of monocrystalline Pt(111) films or monocrystal heated up to temperatures 900-950$^o$C at the pressure ~5x10$^{-8}$ mbar during 60 minutes. This method of synthesis of graphene is self-limited. The reaction stops when the whole surface is covered by monolayer of graphene. Cracking of graphene at Pt(111) leads immediately to formation of quasi-free-standing graphene electronic structure. Other way of formation of the Graphene/Pt interface is intercalation of Pt underneath a graphene synthesized on monocrystalline SiC(0001). Intercalation of Pt monolayer underneath a graphene on SiC can be done by the following operations: firstly Pt is deposited on top of the MG/SiC and than the system is annealed at 600$^o$C during 15 minutes. The (Ni-Fe)-nanodots with intermediate Au monolayer between the graphene and (Ni-Fe) are formed by consequent deposition of Au and Ni, Fe with corresponding concentration on top of Graphene/Pt system (stripes). Intermediate Au monolayer is demanded for restoring the graphene Dirac cone structure which is destroyed under direct contact of graphene with Ni.

Measurements of the photoemission spectra shown in Figs.1 and 2 were carried out at U125/2-SGM and UE112-PGM1 beamlines at BESSY-II (Helmholtz-Zentrum Berlin) by angle- and spin-resolved photoemission at photon energy of 62eV. Part of the experiments was carried out at the Research Resource Center of Saint Petersburg State University "Physical methods of surface investigation".

Modelling results shown in Figs.4 are perfomed by numerical integration of the Landau-Lifshitz-Gilbert equation using our micromagnetic finite-difference code SpinPM based on the forth-order Runge-Kutta method with an adaptive timestep control for the time integration and a mesh size $2\times2\times2$ nm$^3$. For simulation of the Rashba effect the following effective field were added to the LLG equation: $\vec{B}_R = \tilde{\alpha}\left[\vec{j}\times\vec{z}\right]$, where $\vec{j}$ is the in-plane current density vector, $\vec{z}$ is the unite vector along the z-axis direction and $\tilde{\alpha}$ is the experimental spin-orbit constant reported in this work.. The NiFe magnetic parameters used in the modelling are follows: $M_S$=800 emu/cm$^3$, the exchange constant is A= $1.3\times10^{-6}$ erg/cm, Gilbert damping constat is $\alpha^*$ =0.01 and the anozotropy constant is K=$0.5\times10^4$ erg/cm$^2$.

**References**

1. Bauer, G.E.W. et al. Spin caloritronics. *Nature Materials* **11**, 391-399 (2012).
2. Bader, S.D., Parkin, S.S.P. Spintronics. *Annu. Rev. Condens. Matter Phys.* **1**, 71-88 (2010).
3. Awschalom. D. et al. Spintronics without magnetism. *Physics* **2**, 50 (2009).
4. Parkin. S.S.P. et al. Magnetic domain-wall racetrack memory. *Science* **320**, 190-194 (2008).
5. Manchon, A. et al. Theory of nonequolibrium intrinsic spin torque in single nanomagnet. *Phys. Rev. B* **78**, 094422 (2009).
6. Manson, A. et al. Theory of spin torque due to spin-orbit splitting. *Phys. Rev. B* **79**, 212405 (2009).
7. Wang, X. et al. Rashba spin torque in an ultrathin ferromagnetic metal layer. arXiv:1111.5466v1 [cond-mat.mtrl-sci].
8. Miron, I.M. et al. Current-driven spin torque induced by the Rashba effect in a ferromagnetic layer. *Nature Materials* **9**, 230-234 (2010).
9. Miron, I.M. et al. Perpendicular switching of a single ferromagnetic layer induced by in-plane current injection. *Nature* **476**, 189-194 (2011).
10. Emoli, S. et al. Interfacial current-induced torque in Pt/Co/GdOx. Appl. Phys. Lett. **101**, 042405 (2012).
11. Chen, J. et al. Spin torque due to non-uniform Rashba spin orbit effect. AIP Advances **2**, 042133 (2012).
12. Pesin, D.A. et al. Quantum kinetic theory of current-induced toeque in Rashba ferromagnet. *Phys. Rev. B* **86**, 114416 (2012).
13. Liu, L. et al. Spin-torque ferromagnetic resonance induced by the Spin Hall Effect. *Phys. Rev. Lett.* **106**, 036601 (2011).
14. Liu, L. et al. Current-Induced Switching of Perpendicularly Magnetized Magnetic Layers Using Spin Torque from the Spin Hall Effect. *Phys. Rev. Lett.* **109**, 096602 (2012).
15. Liu, L. et al. Magnetic switching by spin torque from the spin Hall effect. arXiv:1110.6846.
16. Novoselov, K.S. et al. Two-dimensional gas of massless Dirac fermions in graphene. *Nature* **438,** 197-200 (2005).
17. Kane, C.L., Mele, E.J. Quantum spin Hall effect in graphene. Phys. Rev. Lett. **95**, 226801 (2005).
18. Shikin, A.M. et al. Origin of spin-orbit splitting for monolayers of Au and Ag on W(110) and Mo(110). *Phys. Rev. Lett.* **100,** 057601 (2008).
19. Varykhalov, A. et al. Quantum cavity for spin due to spin-orbit interaction at a metal bondary. *Phys. Rev. Lett.* **101**, 256601 (2008).
20. Rybkin, A.G. et al. Topology of spin polarization of the 5d states on W(110) and Al/W(110) surfaces. *Phys. Rev. B* **86**, 035117 (2012).
21. Rybkin, A.G. et al. Spin-dependent avoided-crossing effect on quantum-well states in Al/W(110). *Phys. Rev. B* **82**, 233403 (2010).
22. Rybkin, A.G. et al, Large spin-orbit splitting in light quantum films: Al/W(110). *Phys. Rev. B* **85**, 045425 (2012).

23. Marchenko, D. et al. Giant Rashba splitting in graphene due to hybridization with gold. *Nature Commun.* **3**, 1232 (2012).
24. Varykhalov, A. et al. Electronic and magnetic properties of quasifreestanding graphene on Ni. *Phys. Rev. Lett.* **101**, 157601 (2008).
25. Shikin, A.M. et al. Induced spin-orbit splitting in graphene: the role of atomic number of the intercalated metal and π-d hybridization. *New J. Phys.* **15**, 013016 (2013).
26. Kimura, T. et al. Room-temperature reversible spin Hall effect. *Phys. Rev. Lett.* **98**, 156601 (2007).
27. Guo, G.Y. et al. Intrinsic spin Hall effect in Pt: first-principles calculations. *Phys. Rev. Lett.* 100, 096401 (2008).
28. Kontani, H. Et al. Study of intrinsic spin and orbital Hall effect in Pt based on a (6s,6p,5d) tight-binding model. ArXiv:0705.3535v3 [cond-mat.mes-hall] 2008.
29. Emori, S. et al. Current-driven dynamics of chiral ferromagnetic domain walls. *Nature Materials.* **12**, 611-616 (2013).
30. Garello, K. et al. Symmetry and magnitude of spin-orbit torques in ferromagnetic heterostructures. *Nature Nanotechnology* **8,** 587-593 (2013)
31. Ando, K. et al. Electric manipulation of spin relaxation using the spin Hall effect *Phys. Rev. Lett.* **101**, 036601 (2008).
32. Saitoh, E. et al. Conversion of spin current into charge current at room temperature: invese spin Hall effect. Appl. Phys.Lett**. 88**, 182 (2006).
33. Uchida, K. et al. Observation of the spin Seebeck effect. *Nature* **455**, 778-781 (2008).
34. Uchida, K. et al. Spin-Seebeck effect in $Ni_{81}Fe_{19}$/Pt films. *Solid State Commun.* **150**, 524-528 (2010).
35. Popova, A. A. et al. The role of the covalent interaction in the formation of the electronic structure of Au- and Cu-intercalated graphene on Ni(111). Phys. Solid State **53**, 2539 (2011)
36. Rybkina, A.A. et al. The graphene.Au/Ni interface and its application in the construction of a graphehe spin filter. *Nanotechnology* **24**, 295201-295209 (2013).
37. Shikin A.M., et al. Surface intercalation of gold underneath a graphene monolayer on Ni(111) studied by angle-resolved photoemission and hifg-resolution energy-loss spectroscopy. *Phys. Rev. B* **62**, 13202 (2000).
38. D. Dragoman and M. Dragoman. Giant thermoelectric effect in graphene. *Appl. Phys. Lett.* **91**, 203116 (2007).
39. Deacon, R.S. et al. Cyclotron resonance study of the electron and hole velocity in graphene monolayers. *Phys. Rev. B* **76**, 081406(R) (2007).
40. Miron, I.M. et al. Fast current-induced domain-wall motion controlled by the Rashba effect. *Nature Materials* **10**, 419-423 (2011).
41. Karpan, V.M. et al. Graphite and graphene as perfect spin filters. *Phys. Rev. Lett.* **99**, 176602 (2007).
42. Karpan, V.M. et al. Theoretical prediction of perfect spin filtering at interface between close-packed surfaces of Ni or Co and graphite and graphene. *Phys. Rev. B* **78**, 195419 (2008).
43. Tombros, N. et al. Electtonic spin transport and spin precession in single graphene layers at room temperature. *Nature* **448**, 571-574 (2007).
44. Cobas, E. et al. Graphene as a tunnel barrier: graphene-based magnetic tunnel junctions. *Nanoletters* **12**, 3000-3004 (2012).
45. Van't Erve, O.M. et al. Low resistance spins injection into silicon using graphene tunnel barrier. *Nature Nanotechnology* **7**, 737-742 (2012).
46. Zhang, C. et al. Enhancement of spin injection from ferromagnet to graphene with a Cu interfacial layer. *Appl. Phys. Lett.* **101**, 022406 (2012).
47. Ralph, D.C., Stiles, M.D. Spin transfer torques. *J. Magn. Magn. Mater.* **320**, 1190-1216 (2008).

48. Hayashi, M. et al. Current-Controlled Magnetic Domain-Wall Nanowire Shift Register. *Science* **320**, 209-211 (2008).
49. Khvalkovskiy, A.V. et al. High Domain Wall Velocities due to Spin Currents Perpendicular to the Plane. *Phys. Rev. Lett.* **102**, 067206 (2009).
50. Metaxas, P. J. et al. High domain wall velocities via spin transfer torque using vertical current injection. *Scientific Reports* **3**, 1829 (2013).
51. Khvalkovskiy, A.V. et al. Matching domain-wall configuration and spin-orbit torques for efficient domain-wall motion. *Phys. Rev. B* **87**, 020402(R) (2013).
52. Khvalkovskiy, A.V. et al. High Domain Wall Velocities due to Spin Currents Perpendicular to the Plane. *Phys. Rev. Lett.* **102**, 067206 (2009).
53. Metaxas, P. J. et al. High domain wall velocities via spin transfer torque using vertical current injection. *Scientific Reports* **3**, 1829 (2013).
54. Wu, Z.-S. et al. Synthesis of Graphene Sheets with High Electrical Conductivity and Good Thermal Stability by Hydrogen Arc Discharge Exfoliation. *ACS Nano* **3**, 411–417 (2009).

**Acknowledgments**
This work was supported by grant of the Saint Petersburg State University for scientific investigations, RFBR-project no. 11-02-00642-a and DFG-RFBR project no.13-02-91336 (RA 1041/3-1). A.A.R. and I.I.K. acknowledge for support from the German-Russian Interdisciplinary Science Center (G-RISC) funded by the German Federal Foreign Office via the German Academic Exchange Service (DAAD). Authors acknowledge Helmholtz-Zentrum Berlin and Russian-German laboratory at BESSY for the support and the staff of BESSY for the help during the experiment. P.N.S. acknowledges the Dynasty Foundation.

**Author contributions**
A.M.Sh., A.A.R., A.G.R., I.I.K. carried out the experimental measurements and analyzed the data with the discussion of the proposed idea with P.N.S. and A.K.Z. A.K.Z. and P.N.S performed the numerical estimations and mathematical analysis of the proposed idea. P.N.S and K.A.Z. carried out the micromagnetic modeling. A.M.Sh. wrote the paper with assistance of A.A.R., P.N.S. and I.I.K. in the figure preparation and intense discussion.

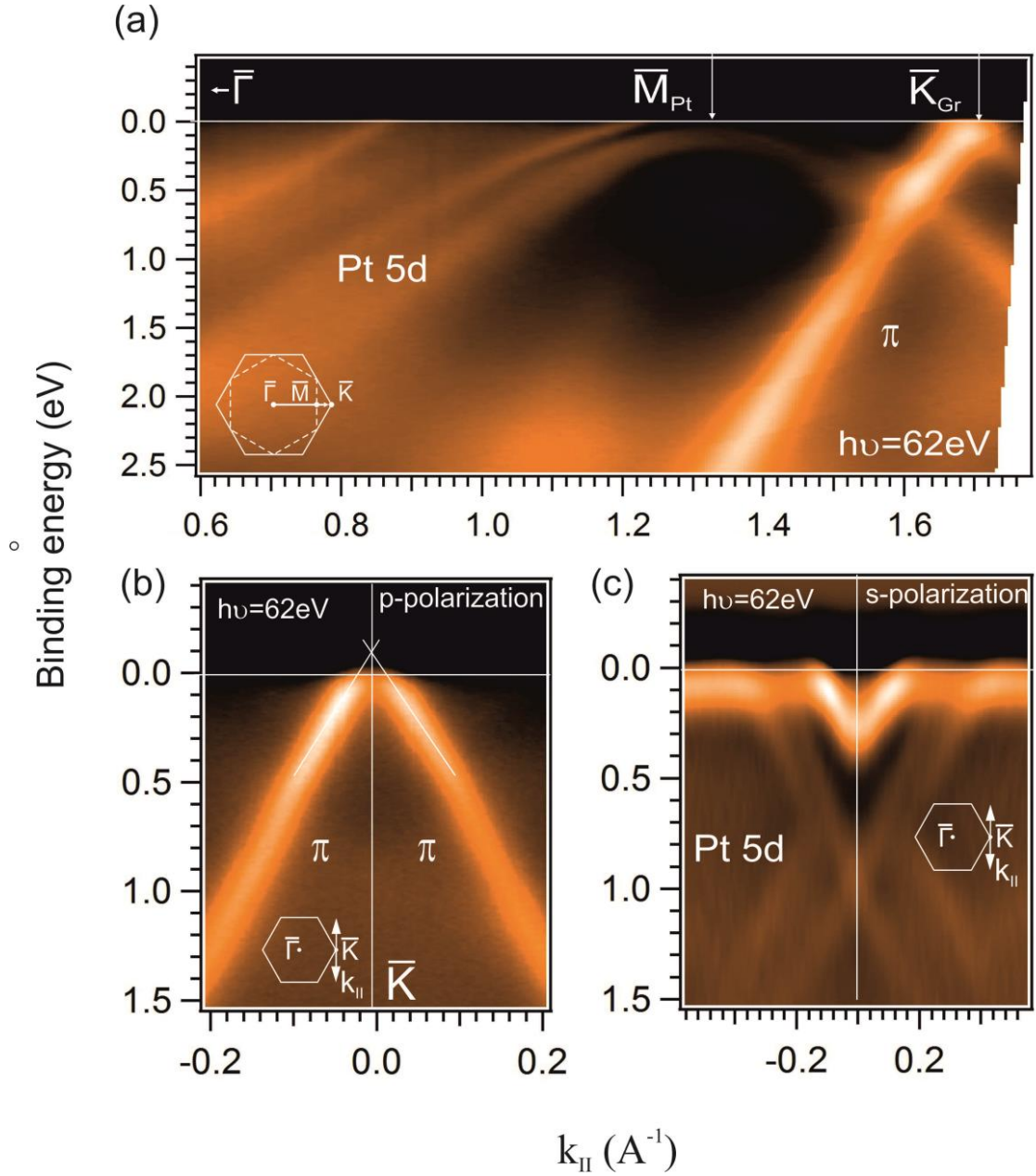

Fig.1. Dispersion relations of the graphene π states and Pt d states for graphene on Pt(111) measured by angle-resolved photoemission in the ΓK direction of the graphene BZ at temperature 18K – (a). The positions of the K-point of the graphene BZ and of the M-point of the Pt(111) BZ are shown. (b) – corresponding detailed dispersion relations for the graphene π states near the K-point of the BZ located at $k_{II}$ = 1.7 Å$^{-1}$ measured in the direction perpendicular to the ΓK. The dispersion diagrams in (a) and (b) were measured with p-polarization of the incident synchrotron radiation. Drowing of the graphene and Pt BZ are shown in insets in Figs 1.a,b by solid and dashed lines respectively. The direction of measurements of the dispersions is shown by arrows. (c) shows the dispersion in the same region as in (b), but for the measurements with use of s-polarization of the incident synchrotron radiation. Such type measurement is insensitive to the π states of graphene and shows mainly the dispersion dependences for the Pt-derived d states.

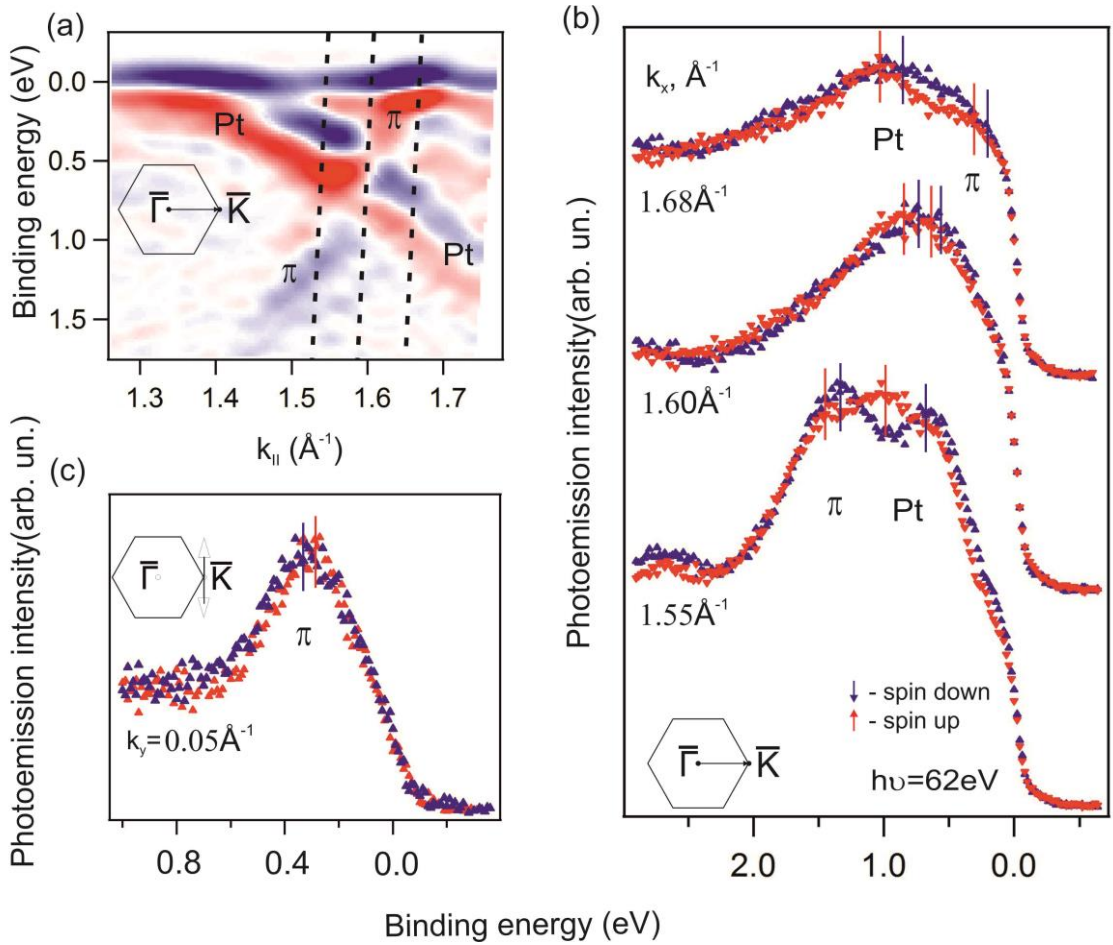

Fig.2. Dispersions of the graphene π and Pt 5d states for the Graphene/Pt interface measured by angle-resolved photoemission with circular polarization of the synchrotron radiation – (a). Blue and red colors correspond to the contributions with opposite (positive and negative) elliptic polarizations. Spin-resolved photoemission spectra for the graphene π states and Pt d states measured in the ΓK direction of the graphene BZ close to the K-point – (b) and perpendicular to the ΓK direction through the K point – (c) demonstrating the spin-orbit splitting of the graphene π and Pt 5d states near the Fermi. The vertical dashed lines in (a) show the values of $k_{II}$ used for measurements of the spin-resolved specra presented in (b).

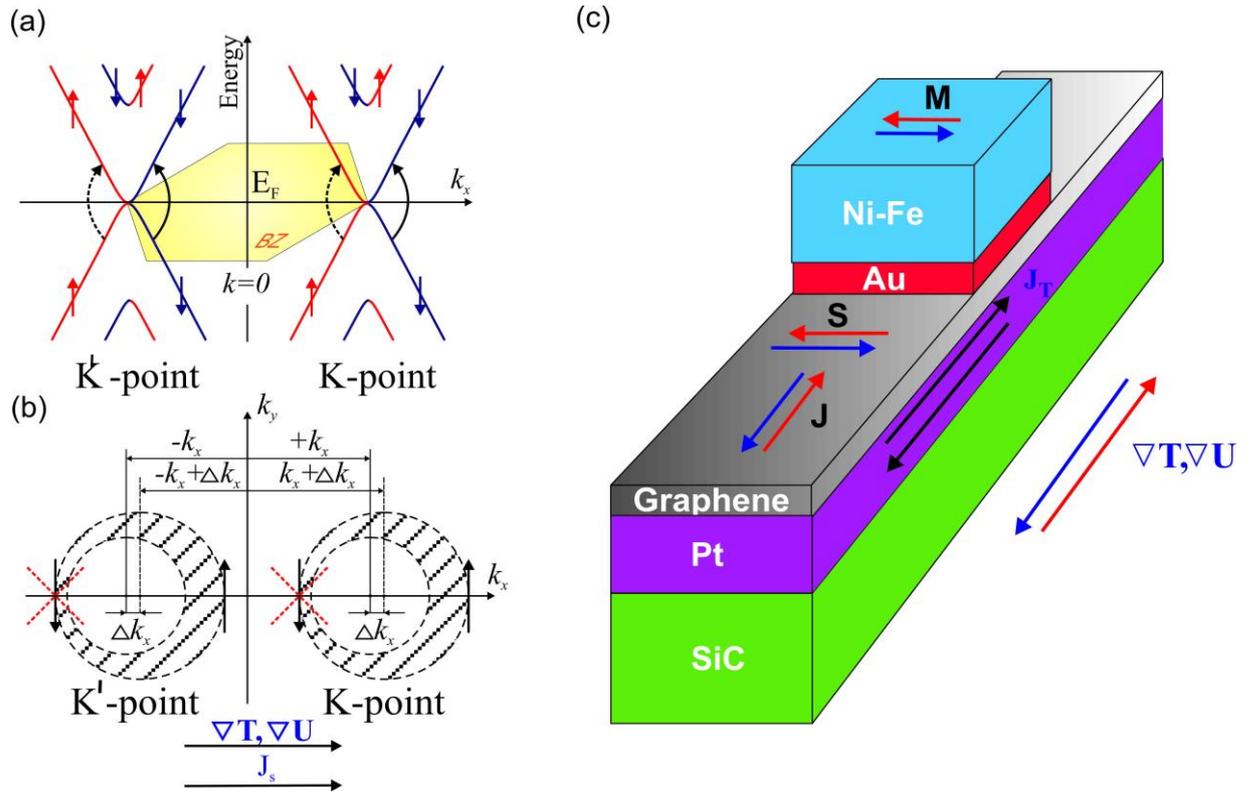

Fig.3. Spin electronic structure of the Dirac-cone graphene π states for the Graphene/Pt interface with a large spin-orbit splitting of the π states at the Fermi lelvel near the K and K' points of the graphene BZ drawn for opposite $k_{II}$ directions - (*a*), and the difference between the contributions to the excited current of electrons with opposite spin orientations realized when setting up an electrical or thermal gradient (∇*T*, ∇U) between the ends of the system - (*b*). Electrons excited into the states characterized by increase in energy with increase of $k_x$ (bent solid arrows in (*a*)) will take part in formation of the spin current in the $k_x$ direction. By contrast, electrons excited into the states characterized by decrease in energy with increase of $k_x$ (dashed bent arrows in (*a*)) will scatter rather than contribute to formation of this current. This is shown by the forbidding red crosses in (*b*). (c) - schematic presentation of the proposed device construction with the (Ni-Fe)-nanodots deposited atop the Graphene/Pt interface. The spin current developed in the Graphene/Pt stripe due to an applied electrical or thermal gradient is characterized by the spin strongly locked perpendicular to the momentum of the moving electrons. It induces a reverse in-plane magnetization in the (Fe-Ni)-nanodots (in dependence on the direction of the spin current) due to the spin-orbit torque effect. The construction with thin Pt-stripes can be arranged on top of SiC.

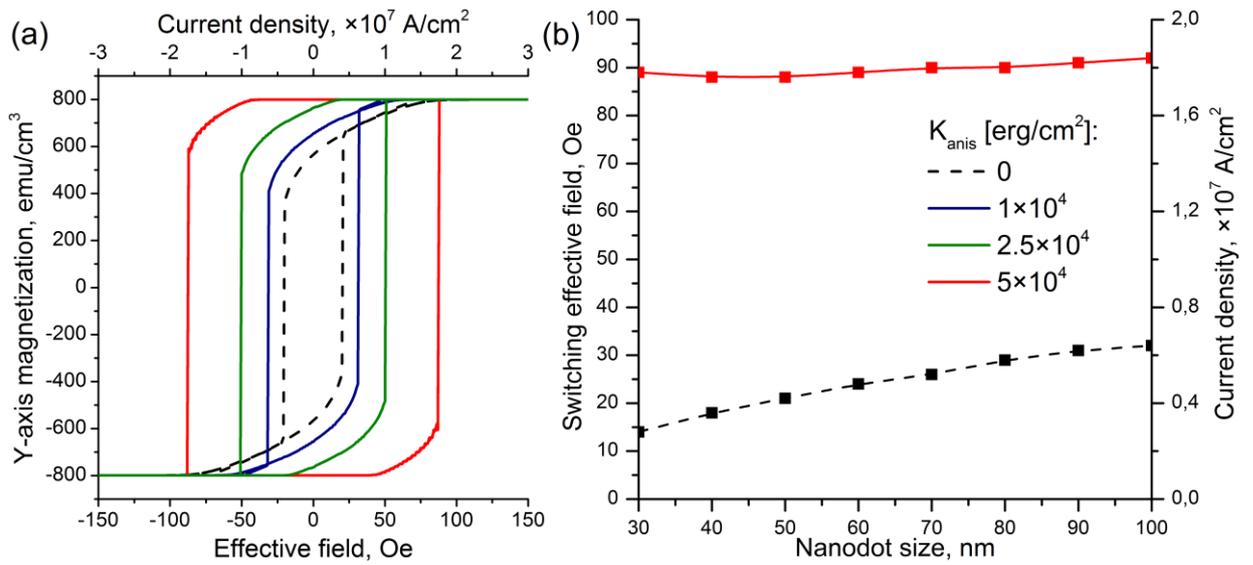

Fig.4. (a) - calculated hysteresis loops for the (Ni-Fe)-nanodot with the sizes $50\times50\times2$ nm$^3$ for different values of the anisotropy constant. (b) - critical remagnetization effective field/current dependens on the nanodot size for hard and soft magnet cases. In the case of a soft magnet (dashed line) the critical switching field is determined by the shape anisotropy. The results show that the width of the hysteresis loop for the NiFe-nanodot is no more than 90 *Oe* or $1.8\times10^7$ A/cm$^2$ for the hard magnets and 30 *Oe* or $6\times10^6$ A/cm$^2$ for the soft magnets, i.e. the (Ni-Fe)-nanodot can be remagnetized by the induced magnetic field formed by the spin current at the Graphene/Pt interface.

# Supplementary information

# Spin structure of Graphene/Pt interface for spin current formation and induced magnetization in deposited (Ni-Fe)-nanodots

A.M. Shikin[1*], A.A. Rybkina[1], A.G. Rybkin[1], I.I. Klimovskikh[1], P.N. Skirdkov[2,3], K.A. Zvezdin[2,3], A.K. Zvezdin[2,3]

[1] Saint Petersburg State University, Saint Petersburg, Peterhof, Ulyanovskaya str. 1, 198504 Russia
[2] A.M. Prokhorov General Physics Institute, Russian Academy of Sciences, Moscow, Vavilova str. 38, 119991 Russia
[3] Moscow Institute of Physics and Technology, Dolgoprudny, Institutskiy per. 9, 141700 Russia

**Spin current developed at the Graphene/Pt interface due to application of an electrical or thermal gradient.**

The process of the spin generation in the graphene-derived system with strong spin-orbit (Rashba) interaction can be described by the following Hamiltonian:

$$H_{SO} = \varepsilon(p)\hat{I} - \frac{\alpha}{\hbar}\left(p_x\sigma_y - p_y\sigma_x\right) \quad (1),$$

where $\varepsilon(p) = \hbar\upsilon k$ is the dispersion law for the electrons of graphene, $\hat{I}$ is the Identity matrix, $\alpha$ is the Rashba constant ($\alpha \cdot k_F = 80$ meV), $\sigma_{x,y}$ are the Pauli matrixes. In the case of the presence of an electrical or temperature gradient we can describe the spin accumulation in the system with strong spin-orbit interaction, using the Boltzmann equation in time-relaxation approximation with additional force $F_x$, with $F_x^{\nabla T} = (\partial T/\partial x)\cdot(\partial\mu/\partial T)$ if it caused by the temperature gradient, where $\mu$ is a chemical potential, and $F_x^{\nabla U} = eE_x$ in case of applied electric field $E_x$. The solution of this equation can be represented in the form:

$$f = f_0 + \upsilon_x F_x \tau \frac{\partial f_0}{\partial \varepsilon} \quad (2),$$

where $\tau$ is the relaxation time, $f$ is the distribution function of electrons, $f_0$ is the unperturbed distribution function of electrons. Application of the thermogradient $\partial T/\partial x$ or electric field $E_x$ along x direction causes a spin current alone the x axis. In our case spin current can be defined as follows:

$$j_S^x = j_+^x - j_-^x = B\frac{e\tau}{\hbar^2}F_x\int d^2\vec{k}\left(\frac{\partial\varepsilon^+}{\partial k_x}\frac{\partial f_0^+}{\partial k_x} - \frac{\partial\varepsilon^-}{\partial k_x}\frac{\partial f_0^-}{\partial k_x}\right) \quad (3),$$

where $B = (n/2\pi)\cdot(\upsilon\hbar/\varepsilon_F)^2$, $n$ is the electron density in graphene, $f_0^\pm$ is the unperturbed distribution function of majority/minority electrons. The following relations can be obtained from Hamiltonian (1):

$$\begin{cases}\frac{\partial\varepsilon^\pm}{\partial k_x} = \hbar\upsilon\left(\frac{k_x}{k}\right) \pm \alpha\left(\frac{k_x}{k}\right) \\ \frac{\partial f_0^\pm}{\partial k_x} = \frac{\partial f_0^\pm}{\partial \varepsilon}(\hbar\upsilon \pm \alpha)\frac{k_x}{k}\end{cases} \quad (4).$$

Using the relations (4), the spin current in case of the thermogradient can be rewritten as:

$$j_S^x = \frac{2\alpha}{\hbar\upsilon} j_e^x = -S\frac{2\alpha}{\hbar\upsilon}\frac{\partial T}{\partial x} \quad (5),$$

where $S = \frac{en\upsilon^2\tau}{\varepsilon_F}\frac{\partial \mu}{\partial T}$ is the Seebeck constant, $j_e^x$ is the electric current along the x-axis. In case of the electrical gradient the spin current can be rewritten as:

$$j_S^x = \frac{2\alpha}{\hbar\upsilon} j_e^x = \frac{2\alpha}{\hbar\upsilon}\sigma E_x \quad (6),$$

where $\sigma$ is the graphene conductivity.

In addition to the spin current, under application of the electric field $E_x$ or the temperature gradient $\partial T/\partial x$ along x direction an uncompensated density spin appears in the y direction (because the spin of electron in the Rashba model is locked perpendicular to the momentum). The spin accumulation $\langle\delta\vec{\sigma}\rangle = B\int d^2\vec{k}\left(f^+ - f^-\right)\vec{e}_y$, where $\vec{e}_y$ is the unit vector along the y-axis, after substitution of (2) can be transformed to the following form:

$$\langle\delta\vec{\sigma}_x\rangle = \langle\delta\vec{\sigma}_z\rangle = 0 \text{ and } \langle\delta\vec{\sigma}\rangle_y = -\alpha\frac{n\tau}{\hbar\varepsilon_F}F_x \quad (7).$$

Substituting the expression of $F_x$ for the cases of thermogradient and applied electric field, we can obtain spin accumulation in following form:

$$\langle\delta\vec{\sigma}\rangle_y^{\nabla T} = -\frac{\alpha}{e\hbar\upsilon^2}S\frac{\partial T}{\partial x} \text{ and } \langle\delta\vec{\sigma}\rangle_y^{\nabla U} = -\alpha\frac{ne\tau}{\hbar\varepsilon_F}E_x \quad (8).$$

**Application of the Graphene/Pt interface for the magnetization dynamics of the (Ni-Fe)-nanodots array due to spin-orbit torque effect.**

Spin current in Graphene/Pt plane enters to the NiFe contact and affects the effective magnetic field in NiFe. Effective Hamiltonian in this case can be described as:

$$H = \varepsilon(p) + H_R - J\vec{M}^0\langle\vec{\sigma}\rangle \quad (9),$$

where $\vec{M}^0 = \vec{M}/M_S$ is the unit vector along the magnetization direction. The last term is related to the induced magnetic field, $\langle\vec{\sigma}\rangle$ is the uncompensated spin, $M_S$ is the saturation magnetization of magnetic material, $J$ is the exchange integral. The second (Rashba) term $H_R = \alpha\left[\vec{k}\times\vec{z}_0\right]\vec{\sigma}$ leads to the appearance of the effective magnetic field $\vec{B}_{SO}$:

$$\vec{B}_{SO} = -\frac{\alpha k_F}{e\upsilon M_S}P\cdot j_e\left[\vec{z}^0\times\vec{j}^0\right], \text{ where } j^0 = 1, P = \frac{J}{\varepsilon_F} \quad (10).$$

This additional effective field leads to appearance of the spin torque, acting on magnetization in NiFe layer. Corresponding spin torque per unit volume is given by:

$$\vec{T}_{SO} = \left[\vec{M}\times\vec{B}_{SO}\right] = \frac{\alpha k_F}{e\upsilon M_S}Pj_e\left[\vec{M}\times\vec{j}^0\right] = \frac{\alpha k_F}{e\upsilon}Pj_e\left[\vec{M}^0\times\vec{j}^0\right] \quad (11).$$